\documentclass[11pt]{article}

\usepackage{amsmath}
\usepackage{amsfonts}
\usepackage{color}
\usepackage[procnames]{listings}
\usepackage{textcomp}
\usepackage{setspace}
\usepackage{palatino}
\usepackage{graphicx}
\usepackage{cite}
\usepackage{url}
\usepackage[utf8]{inputenc}
\usepackage[T1]{fontenc}
\usepackage[french]{babel}

\definecolor{gray}{gray}{0.5}
\definecolor{green}{rgb}{0,0.5,0}
\definecolor{lightgreen}{rgb}{0,0.7,0}
\definecolor{purple}{rgb}{0.5,0,0.5}
\definecolor{darkred}{rgb}{0.5,0,0}

\makeatletter
\def\@sect#1#2#3#4#5#6[#7]#8{%
  \ifnum #2>\c@secnumdepth
    \let\@svsec\@empty
  \else
    \refstepcounter{#1}%
    \protected@edef\@svsec{\@seccntformat{#1}\relax}%
  \fi
  \@tempskipa #5\relax
  \ifdim \@tempskipa>\z@
    \begingroup
      #6{%
        \@hangfrom{\hskip #3\relax\@svsec}%
          \interlinepenalty \@M #8\@@par}%
    \endgroup
    \csname #1mark\endcsname{#7}%
    \addcontentsline{toc}{#1}{%
      \ifnum #2>\c@secnumdepth \else
        \protect\numberline{\csname the#1\endcsname.}%
      \fi
      #7}%
  \else
    \def\@svsechd{%
      #6{\hskip #3\relax
      \@svsec #8}%
      \csname #1mark\endcsname{#7}%
      \addcontentsline{toc}{#1}{%
        \ifnum #2>\c@secnumdepth \else
          \protect\numberline{\csname the#1\endcsname.}%
        \fi
        #7}}%
  \fi
  \@xsect{#5}}
\def\@seccntformat#1{\csname the#1\endcsname.\quad}
\makeatother

\makeatletter

\@addtoreset{equation}{section}
\makeatother

\newcommand{\carrenoir}{\rule{0.5em}{0.5em}}


\newcommand{\oper}[2]{\newcommand{#1}{\mathop{\mathrm{#2}}\nolimits} }
\oper{\Vol}{Vol}

\oper{\Ker}{Ker}
\oper{\Ima}{Im}
\oper{\dimension}{dim}

\oper{\injrad}{inj}

\DeclareSymbolFont{greek}{OML}{ptmcm}{m}{it}
\DeclareMathSymbol{\codiff}{\mathord}{greek}{"0E}
\DeclareMathSymbol{\prodint}{\mathord}{greek}{"13}

\title{Interaction entre math\'ematique et informatique par le logiciel math\'ematique Libre/Open Source}
\author{K.I.A.Derouiche}
\date{}
\begin{document}

\maketitle
{\small 
\textsc{Résumé.---}
Cet article porte sur l'application du mod\`ele de d\'eveloppement et 
l'ouverture du code source, disponible et mis en oeuvre par les Logiciels Libres
 et Open Sources (LL/OS) \`a la didactique et l'enseignement a la fois des 
math\'ematiques et l'informatique par la lecture-\'ecriture (L/E) de logiciel 
math\'ematique, dont les cas les plus connus sont le calcul num\'erique et 
formel.

L'article analyse le mod\`ele de d\'eveloppement du logiciel math\'ematique 
Libre /Open Source (L/OS) dont l'importance est av\'er\'ee dans le secteur
de la recherche en math\'ematique et informatique. En revanche, leur 
utilisation bien que r\'eelle, est peu lisible dans les formations 
d'enseignement sup\'erieur. Nous discutons de la faisabilit\'e de ce 
mod\`ele, qui concerne les caract\'eristiques du domaine, des acteurs,
des interactions qu'ils entretiennent et des communaut\'es qu'ils forment
pendant le d\'eveloppement de ces logiciels. Finalement, on propose un exemple
de logiciel math\'ematique Libre/Open Source  (LML/OS) comme dispositif
d'analyse.

Mots-clefs : Enseignement LL/OSS, Open Source, Coop\'eration, logiciel 
math\'ematique, didactique de l'informatique.

(S\'eminaire National sur la didactique des Math\'ematiques 25-26 Novembre 
Tebessa, Alg\'erie (SNDM'13))}

\section{Introduction}
Après avoir été un outil réservé aux centres de recherches. 
LL/OS s'est implanté dans l'industrie, et depuis quelque années, 
il est omniprésent dans pratiquement tous les secteurs d'activités 
de la vie quotidienne, notamment dans  les domaines de la
gestion, de l'industrie, des sciences et techniques et l'éducation 
dont il sont souvent issue. Leur principale innovation concerne leur 
modèle de développement coopératif et décentralisé qui s'est avérée,
dans plusieurs des cas\cite{OO12}, 
plus efficace que le modèle propriétaire.

Notre objectif vise à montrer l'apport des LML/OS dans l'enseignement et
l'apprentissage des mathématiques, qui nécessitent. la qualification des 
enseignants, le savoir acquis de la part des étudiants, une dynamique de
coopération, partage de connaissances, la participation et la discussion,
l'adaptation des contenus pédagogiques mathématiques à l'apprentissages. 
D’où, notre intérêt envers le modèle de développement des LML/OS dans 
l'enseignement et l'apprentissage des mathématiques, que nous allons 
aborder à travers les questions suivantes :
\begin{enumerate}
  \item Quels logiciels mathématiques L/OS utiliser ?
  \item Quels impacts des modifications/contributions des LML/OS sur
l'enseignement et l'apprentissage des mathématiques ?
\end{enumerate}

\section{D\'efinitions de Logiciels Libres/Open source}
\label{defllos}
Il existe plusieurs définitions du ''logiciel libre''. Voir \cite{YT06,UUS05} 
de comparaison entre open source, free software et des divers sens 
qu'on peut donner au LL/OS. Ce qui importe dans notre étude, c'est à la 
fois un outil et modèle de développement. Nous nous
plaçons dans la définition donnée par la Free Software Fondation\cite{FS13} 
qui fait référence à la liberté pour les utilisateurs d'exécuter, 
de copier, de distribuer, d'étudier, de modifier et d’améliorer le logiciel.
 Plus précisément, elle fait référence à quatre
types de libertés:
\begin{enumerate}
  \item La libert\'e d'ex\'ecuter le programme, pour tous les usages (liberté 0)
  \item La libert\'e d'\'etudier le fonctionnement du programme, 
et de l'adapter \`a vos besoins (liberté 1). Pour ceci 
l'acc\'es au code source est une condition
requise
   \item La libert\'e de redistribuer des copies, donc d'aider 
votre voisin, (libert\'e 2)
   \item La libert\'e d'am\'eliorer le programme et de publier vos 
am\'eliorations, pour en faire profiter toute la communauté (libert\'e 3). 
Pour ceci l'accès au code source est une condition requise
Un programme est un logiciel libre si les utilisateurs. 
ont toutes les libertés 0, 1, 2, 3.
\end{enumerate}
\section{Le mod\`ele de d\'eveloppement}
\label{devel}
Le modèle de développement de LL/OS est apparu comme une alternative au modèle
propriétaire de développement du logiciel. Il propose une approche nouvelle de
collaboration entre acteurs (développeurs, testeurs, mainteneurs, 
utilisateurs avancées newbies)\cite{HH11}, de l'utilisation des droits 
de propriété intellectuelle et de la dynamique interne du processus (outils de 
communications, outils de développements, code source)\cite{FB06} Cette 
approche alternative serait à la source du succès de plusieurs LL/OS, 
comme les logiciels mathématiques.
\subsection{Le modèle du logiciel propriétaire}
Le mod\`ele de logiciel propri\'etaire, ou ''modèle d'investissement 
privé''\cite{VH05}, constitue la manière la plus répandue de création 
de technologie, associée aux grandes découvertes technologiques 
qui support la croissance économique depuis un siècle et demi. 
Son idée de base et d'encourager l'innovation par la création de monopoles
 temporaires protégés par les mécanismes de protection de la propriété 
intellectuelle, qui excluent les non créateurs d'une nouvelle technologie 
de son utilisation. Ce modèle de développement ne permet pas un contr\^ole 
sur le logiciel, seulement dans le cadre d'une utilisation limitée par la 
licence
\pagebreak
\begin{figure}[!t]
  \centering
  \includegraphics[scale=0.5]{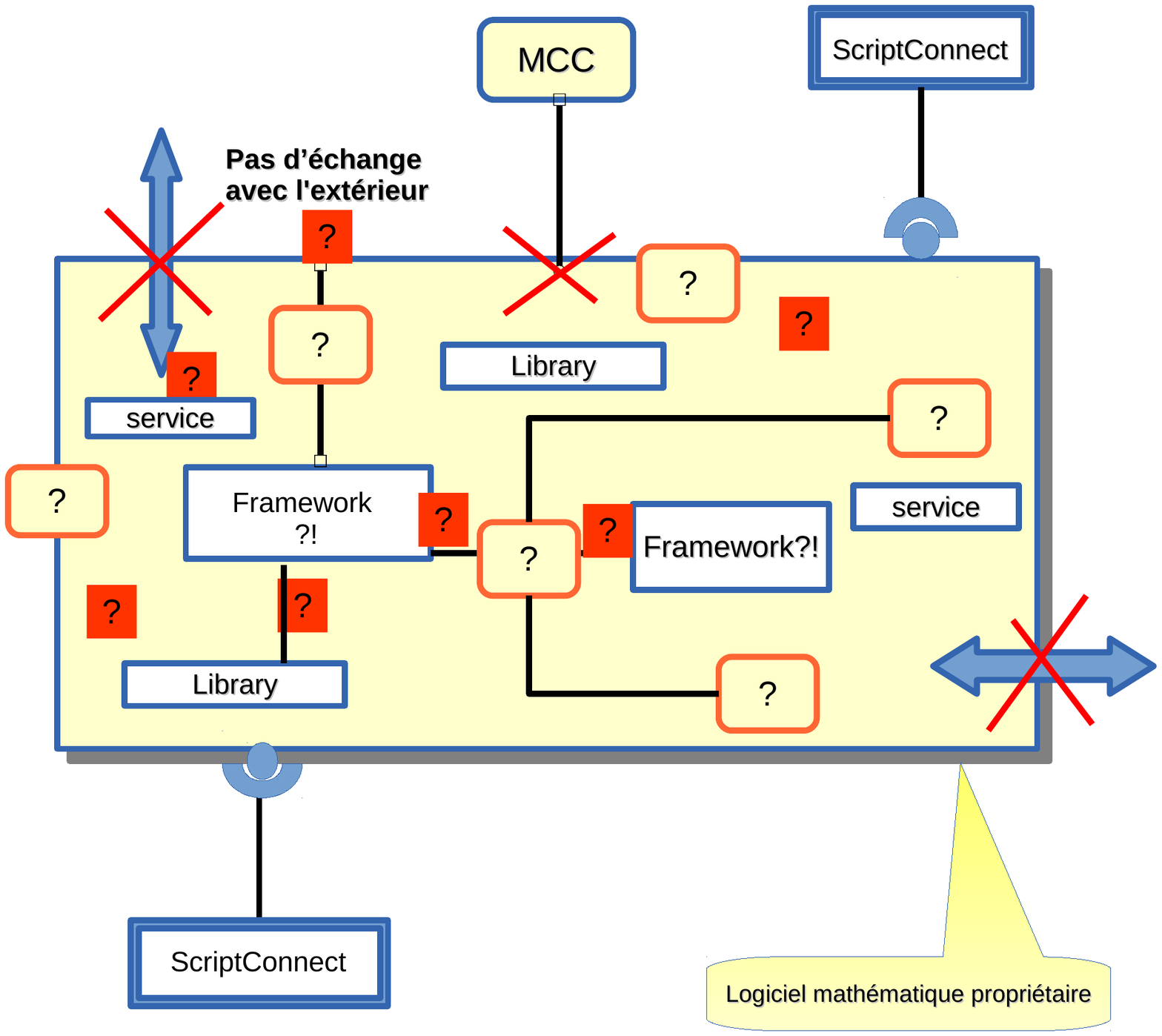}
  \caption{Mod\`ele d'interaction des logiciels math\'ematiques 
propri\'etaires.}
  \label{fig:dia1}
\end{figure}
Un cas pratique est le besoin de faire des corrections après avoir 
d\'etect\'e une erreur (figure \ref{fig:dia1}). Pèrez et Varona\cite{PV13} 
ont \'et\'e confronté à une erreur de calcul de matrices de nombres 
entiers dans le logiciel  Mathematica, qui donne des résultats différents 
s'il évalue le m\^eme déterminant à deux reprises, leur démarche été 
d'isoler l'erreur par un ScriptConnect, par manque de code source qui 
leur aurait permis d'apporter une correction par le biais du code métier 
MathematicalCodeComponent(MCC)
\subsection{Le modèle LL/OS: le logiciel mathématique}
\label{sec:llos}
Une caract\'eristique distinctive de ces logiciels dans 
l'enseignement des math\'ematiques, est l'encouragement 
\`a la motivation individuelle et collective.
Il vise à développer
\begin{figure}[ht]
 \centering
  \includegraphics[width=9cm]{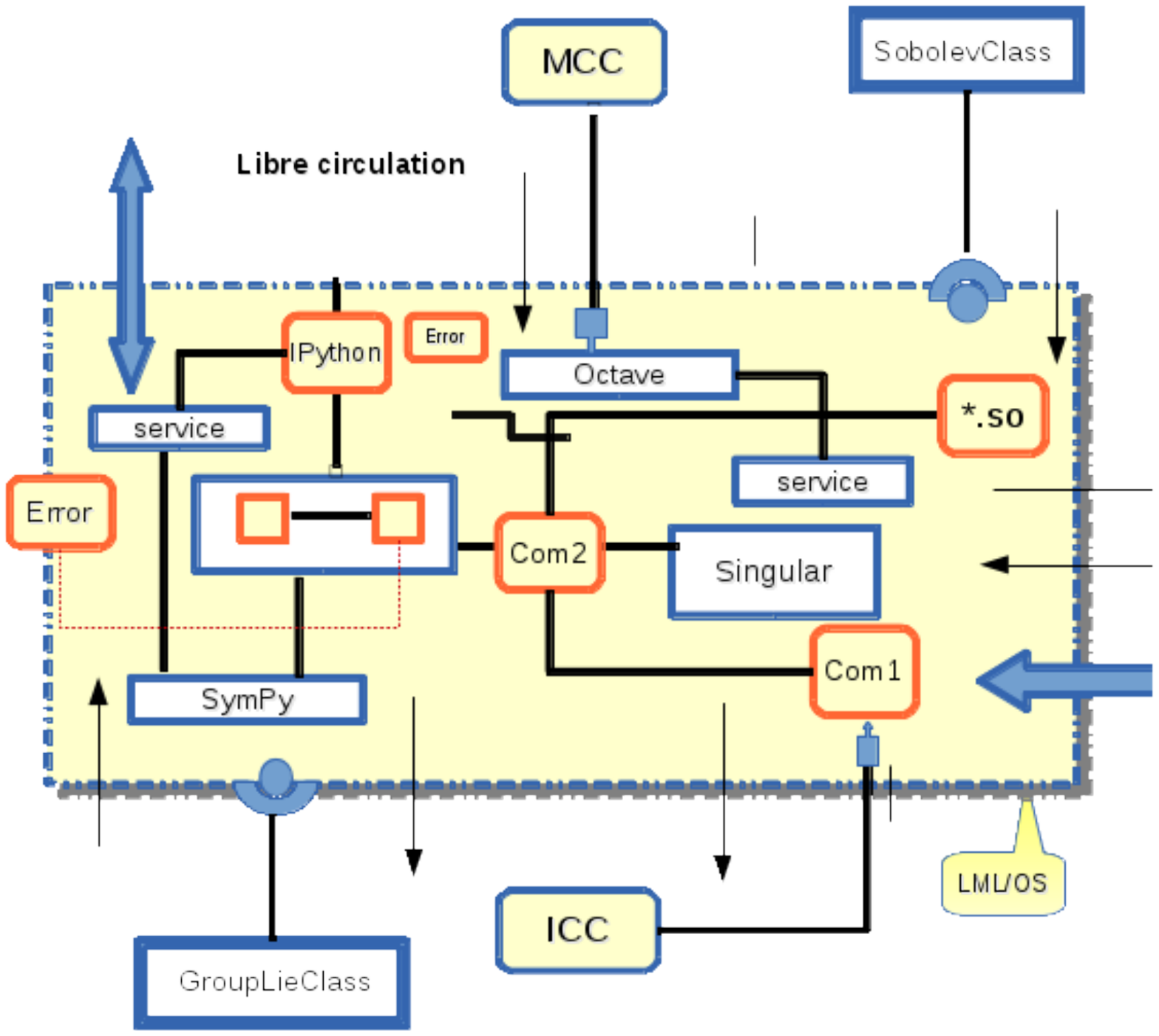}
  \caption{ Mod\`ele d'interaction des LML/OS}
  \label{fig:dia2}
\end{figure}
chez les étudiants, conjointement et progressivement les
capacités d'expérimentation et de raisonnement (exemple: imagination, analyse
critique à travers une démarche de résolution de problèmes, modélisation des
situations et d'apprentissage progressif de la démonstration).
Ils peuvent prendre conscience de la pertinence des activités mathématiques,
identifier un bug (informatique, dysfonctionnement d'un algorithme 
mathématique), le corriger et le redistribuer, l'améliorer et l'expérimenter 
sur des exemples et changé avec l'environnement de travail sans aucune 
restriction voir (figure \ref{fig:dia2})

\section{Quelques logiciels math\'ematiques L/OS}
On peut classer les logiciels math\'ematiques L/OS d\'edi\'es \`a l'enseignementselon six cat\'egories:
\begin{enumerate}
   \item Logiciel biblioth\`eque: gsl, blas, lapack, cgal, sympy, networkx
   \item Logiciel interactif: IPython
   \item Logiciel interface graphique: Cantor
   \item Logiciel Environnement de d\'eveloppement int\'egr\'e: Aesthete, 
Spyder, Reinteract
   \item Logiciel on-web: wims, , sagemath, SymPyLive
\end{enumerate}

\section{Exemple de SymPy}

SymPy\cite{SY13} est une bibliothèque pour les mathématiques symboliques et
 l'algèbre informatique, écrite en Python pure sans dépendance externe, 
Le projet contient trois outils un interpréteur intégré (isympy) 
est deux interface web SymPy Live et SymPy Gamma qui permettent 
aux utilisateurs d'utilisés du code scientifique en ligne. La
bibliothèque SymPy offre d'autres fonctionnalités telles 
que LaTeX, la cryptographie et la mécanique et 
l'informatique quantique, géométrie différentielle.
\lstset{frame=single}
\begin{lstlisting}[language=Python, 
                tabsize=1,
                ]
IPython console for SymPy 0.7.4 (Python 2.7.5-64-bit)
 (ground types: python)

These commands were executed:
>>> from __future__ import division
>>> from sympy import *
>>> x, y, z, t = symbols('x y z t')
>>> k, m, n = symbols('k m n', integer=True)
>>> f, g, h = symbols('f g h', cls=Function)

Documentation can be found at http://www.sympy.org


\end{lstlisting}

\subsection{Expérience de modification/contribution}
\label{sec:exp}
Cette expérience constitue un exemple d'interaction entre mathématique et
informatique dans un environnement de travail coopératif basé sur le modèle du
LL/OS.
L’expérience consiste calculer la dérivée de la fonction $\cos(3x)$
 avec la notation df en utilisation la bibliothèque de calcul symbolique
 SymPy. Pour la mise en oeuvre pratique, nous avons divisé la l'expérience 
en deux tâche mathématique (m\_t\^ache) et une tâche informatique (i\_t\^ache)

\begin{itemize}
  \item \textbf{T\^ache mathématique}
    m\_t\^ache: statbilisation du milieu
    Les \'etudiants doivent:
    \begin{enumerate}
      \item Faire la différence entre $\mathrm{df}$ et $\mathrm{diff}$
      \item Calculer la dérivée de la fonction $f(x)=\cos(3x)$, avec $x \in R$ 
en utilisant la notation de Leibniz
    \end{enumerate}
  \item \textbf{T\^ache informatique}
    \begin{enumerate}
       \item i\_t\^ache0: \'ecrire lexpression symbolique de $\cos(3x)$
         dans isympy en utilisant la notation de Leibniz $\mathrm{df}$
       \item  i\_t\^ache1: comprendre le message d'erreur, localis\'e le 
         fichier source et la fonction qui l'impl\'emente
       \item  i\_t\^ache2: changer le nom de la fonction $\mathrm{diff}$ en 
         $\mathrm{df}$, puis exécuter
    \end{enumerate}
\end{itemize}

\begin{enumerate}
 \item \textbf{Dimension didactique: }
   Pour l'enseignant, il s'agit d'apprendre à construire et \`a conduire des 
   groupes d'\'etudiants dans un environnement de travail coopératif issu 
   du LL/OS. Pour les étudiants, il s'agit d'apprendre \`a construire des 
   connaissances significatives de calcul de d\'eriv\'ee $\cos(3x)$ \`a travers 
   d'une part, les interactions entre le registre symbolique de l'écriture 
   mathématique et le registre historique d'autre part, à travers des tâches 
   à réaliser alternativement dans un environnement de travail coopératif LL/OS.
 \item m\_t\^ache: stabilit\'e du milieu
 Durant toute la tâche, les étudiants travaillent en papier/crayon. Deux 
 objectifs sont particulièrement travaillés:
 \begin{enumerate}
   \item Comprendre l'origine de la notation df (recours à l'histoire des 
  mathématiques et a l’enseignent)
   \item Calculer la dérivée de $\cos(3x)$
 \end{enumerate}
    \textbf{Solution :}
   Explication et discussion de l'enseignent
   Application immédiat du théorème de la dérivée des fonctions trigonométriques
  \begin{displaymath} cos(3x)= -3sin(3x) \end{displaymath}
 \item i\_t\^ache0 : Dans cette t\^ache, les \'etudiants utilisent 
l'invite de commande isympy
     \begin{enumerate}
       \item  Calculer la dériv\'ee de la fonction sous forme symbolique dans
isympy en utilisant $\mathrm{diff}$
     \end{enumerate}
Première tentative: certains étudiants ont appliqué littéralement l'expression
mathématique $\cos(3x)$, et ont \'et\'e surpris du message d'erreur affiché 
dans isympy.
\lstset{frame=single}
\begin{lstlisting}[language=Python, 
                tabsize=1,
                ]
In [1]:diff(cos(3x), x)
  File "<ipython-input-4-762bbb5de069>", line 1
    diff (cos(3x), x)
               ^
SyntaxError: invalid syntax

\end{lstlisting}

\textbf{Solution attendu}:
les étudiants lance une recherche documentaire, dans isympy, en tapent help(cos)
\lstset{frame=single}
\begin{lstlisting}[language=Python, 
                tabsize=1,
                ]
In [2]:diff(cos(3*x), x)
In [2]:-3.sin(3.x)
\end{lstlisting}
\begin{enumerate}
 \item Calculé la dérivée de la fonction cos(3x) en utilisant la notation de
Leibniz $df$ L'enseignant pousse l'étudiant à exploré d'autre possibilité 
pour écrire la dérivée de $\cos (3x)$
\end{enumerate}
\lstset{frame=single}
\begin{lstlisting}[language=Python, 
                tabsize=1,
                ]
In [6]:df(cos(3*x), x)
----------------------------------------------
NameError  Traceback (most recent call last)
<ipython-input-6-9a8e0302dd96> in <module>()
----> 1 df (cos(3*x), x)

NameError: name 'df' is not defined

\end{lstlisting}
i\_t\^ache1: Analyser le message d'erreur, localiser le fichier source de la
fonction qui implémente $\mathrm{diff}$
\lstset{frame=single}
\begin{lstlisting}[language=Python, 
                tabsize=1,
                ]
In [8]: cos(2*x).df(x)
--------------------------------------------------
AttributeError  Traceback (most recent call last)
<ipython-input-8-b449e4a43e07> in <module>()
----> 1 cos(2*x).df(x)

AttributeError: 'cos' object has no attribute 'df'
\end{lstlisting}
La réaction des étudiants face a cette erreur est divisée entre deux groupes, 
ceux qui sont surpris du résultat pour eux les LML/OS c'est avant tout quelque 
chose de très flexible cela implique que il doit être dynamique et immédiat, 
l'autre groupe essai de comprendre la cause des erreurs affichées sur l'écran 
\textbf{AttributeError:'sin object has no attribute 'df' et NameError: name}
\textbf{'diff' is not defined}. L’enseignant explique aux étudiants 
la différence qui existe entre les deux messages d'erreurs.

\textbf{Localisation:}
L'enseignant demande aux étudiants de lancer une commande, pour trouver les
sources de la bibliothèque SymPy dans le système d'exploitation: fournissent 
à la commande le nom du répertoire et un mot clef sur fonction obtenue par 
le biais de la commande  d'aide  help(diff)  dans  l'invite  isympy,
sympy/mpmath/calculus/differentiation.py

i\_t\^ache2: Changer le nom de la fonction dans le fichier source 
de $\mathrm{diff}$ en $\mathrm{df}$, puis tester avec isympy Le r\^ole 
de l'enseignant est  d'impliquer les étudiants dans la L/E approfondie 
du code source et de découvrir tout les enjeux. Dans cette t\^ache 
l'enseignant demande aux étudiant de travailler en collaboration 
sur la L/E du fichier source.

\textbf{Code source Avant les modifications :}
\lstset{frame=single}
\begin{lstlisting}[language=Python, 
                tabsize=1,
                ]
$./sympy/mpmath/calculus/differentiation.py$
....
def df(ctx, f, x, n=1, **options):
"""
...
"""
  partial = False
    try:
        orders = list(n)
        x = list(x)
        partial = True
    except TypeError:
        pass
...
\end{lstlisting}

\textbf{Code source apr\'es les modifications}
\lstset{frame=single}
\begin{lstlisting}[language=Python, 
                tabsize=1,
                ]
...
def df(ctx, f, x, n=1, **options):
...
\end{lstlisting}

\textbf{Ex\'ecution du calcul}
\lstset{frame=single}
\begin{lstlisting}[language=Python, 
                tabsize=1,
                ]
In [30]:df(cos(3*x), x)
In [30]:-3.sin(3.x)
\end{lstlisting}
\end{enumerate}

\section{ Quelles leçons peut-on tirer de cette expérience}
L'utilisation des LML/OS s’avèrent un moyen intéressant pour mettre en oeuvre 
une véritable activité mathématique:

\begin{flushleft}
  \textbf{Avantages}
    \begin{itemize}
 \item  Les étudiants perçoivent les problèmes mathématiques autrement, tout en
respectant un niveau de rigueur. Leur attitude face aux mathématiques
évolue positivement.
\item Aux étudiants de reprendre plusieurs fois un algorithme mathématique, 
ou de revenir plus tard sur certains fragment.
\item La progressivité des  difficultés dans les codes sources qui traites
le problème mathématique, ce qui constitue un réel atout de motivation
pour l'étudiant en difficulté.
\item L'utilisation de ces logiciels est exposée à des effets réseaux 
positifs relier différents cadres (algébrique, géométrique, programmation, 
correction de bug ...) d'un même concept ou d'une  même situation.
\item Les étudiants: peuvent juger, avec relative facilité, la viabilité
 des programmes et algorithmes mathématiques de leur camarade
\item De procéder rapidement à la vérification de certains résultats obtenus.
\end{itemize}
\textbf{Inconv\'enients}
\begin{itemize}
\item  L'étudiant se préoccupe le plus souvent par l'amélioration du 
code source que par le problème mathématique lui-même.
\item Ces logiciels nécessitent parfois la maitrise de divers langage de
programmation, ayant des syntaxes différentes.
\item  Il y a un manque de formation chez les enseignants.
\end{itemize}
\end{flushleft}
\section{Conclusion}
L'interaction entre informatique et mathématique à travers 
l'utilisation des LML/OS dans l'enseignement et l'apprentissage 
des mathématiques s'inscrit dans le champ (discussion, évaluation 
par l'étudiant, travail coopératif, partage de contenus, recherche 
sur des documents et sur le web...). Les LML/OS offrent des opportunités
intéressantes d'exploration dans des situations variées pour 
l'enseignant et pour l'étudiant en l'amenant à réfléchir sur 
ce qu'il fait et comment il doit le faire et avec quel moyen. 
Cependant, nous avons signal\'e un certain nombre d'obstacles qui
peuvent entraver l'utilisation efficace des LML/OS comme. 
la maitrise de la lecture du code source, la maintenance de logiciel, 
l'exécution dans de nouvelle architecture matériel, le facteur 
li\'es à la gestion du temps, le manque de documentation en 
français ou en arabe pour certain logiciels. Les sources de ces difficultés 
sont le manque de formation des enseignants dans ce domaine, 
voire pour les enseignants, le manque de temps(emploi du temps 
chargé, le manque d’intérêts et de volonté de la
part des étudiants ainsi que l'absence d'une véritable 
politique encourageant l'utilisation des logiciels libres 
dans d'apprentissage et d'enseignement des mathématiques.

\bibliographystyle{plain}
\nocite{*}
\bibliography{refloss}

\noindent Kamel Ibn Aziz \textsc{Derouiche}\\
Algerian IT Security Group \\
Cyber Park Sidi Abdellah \\
Incubateur Techno-bridge \\
Route Nationale n°63 \\
Rahmania, Zeralada \\
\texttt{kamel.derouiche@gmail.com}
\end{document}